\documentclass{iau}
\usepackage{graphicx}

\title[Asteroseismology with SuperWASP] 
{Asteroseismology with SuperWASP}

\author[Barry Smalley]   
{Barry Smalley}

\affiliation{Astrophysics Group, Keele University, Staffordshire ST5 5BG, United
Kingdom \\
email: {\tt b.smalley@keele.ac.uk}}

\pubyear{2013}
\volume{301}  
\pagerange{xxx-yyy}
\jname{Precision Asteroseismology}
\editors{Bill Chaplin,
Joyce Guzik,
Gerald Handler, 
\& Andrzej Pigulski, eds.}

\begin{document}

\maketitle

\begin{abstract}
The highly successful SuperWASP planetary transit finding programme has surveyed
a large fraction of both the northern and southern skies. There now exists in
the its archive over 420 billion photometric measurements for more than 31
million stars. SuperWASP provides good quality photometry with a precision
exceeding 1\% per observation in the approximate magnitude range $9 < V < 12$.
The archive enables long-baseline, high-cadence studies of stellar variability
to be undertaken. An overview of the SuperWASP project is presented, along
with results which demonstrate the survey's asteroseismic capabilities.
\keywords{asteroseismology, 
instrumentation: photometers,
stars: chemically peculiar,
stars: oscillations,
stars: rotation,
stars: variables: delta Scuti,
stars: variables: roAp}
\end{abstract}

\firstsection 
\section{Introduction}

The Wide Angle Search for Planets (WASP) is one of the world's leading
ground-based surveys for transiting exoplanets (\cite[Pollacco et al.
2006]{2006PASP..118.1407P}). The project has two robotic telescopes in roll-off
roof enclosures, one  at the Observatorio del Roque de los Muchachos on the
island of La Palma in the Canary Islands, and the other at the Sutherland
Station, South African Astronomical Observatory (SAAO). Both instruments consist
of an array of eight 200-mm, f/1.8 Canon telephoto lenses and 2048$\times$2048
Andor CCDs, provide a field of view of $7.8^{\circ} \times 7.8^{\circ}$ and pixel
size of 13.7$''$. A broad-band filter gives a 400--700~nm bandpass.

SuperWASP observes a set of pre-determined `planet fields' each night,  subject
to their visibility and Moon avoidance. At each pointing, two sequential
30-second exposures are taken. The cameras return to same field with a typical
cadence of around 10 minutes, but this is variable. Typically, in a given
observing season, some 3000 points per star are taken over a period of 100 to
150 days.

The SuperWASP automated photometry extraction pipeline (\cite[Collier Cameron et
al. 2006]{2006MNRAS.373..799C}) uses the Tycho-2 and USNO-B1.0 catalogues to
determine an astrometric solution for each field. Photometry is performed on all
objects, using three different sized apertures, with a 3.5-pixel (48$''$)
aperture being the default used in lightcurves. Given the size of the apertures,
one has to be aware of object blending and dilution. The photometry is
transformed to Tycho-2 $V$ using around 100 stars per field, yielding WASP
pseudo $V$ magnitudes. Lightcurves are sent to the project archive in Leicester.
Further trend removal is performed using the {\sc SysRem} algorithm (\cite[Tamuz
et al. 2005]{2005MNRAS.356.1466T}), which is effective at removing some of the
correlated `red noise' from the lightcurves (\cite[Smith et al.
2006]{2006MNRAS.373.1151S}). Overall photometric performance is better than 1\%
for V $<$ 11.5 and 0.5\% for V$<$9.4.

The SuperWASP archive currently holds over 420 billion data points covering 31
million stars, obtained from 12 million images taken during 2100 nights since
2004. The coverage of the survey is now virtually the entire sky, with the
exception of the Galactic plane where the stellar density is too high to permit
useful aperture-photometry of objects on account of the instrument's large pixel
size. This multi-season and multi-site photometry is an excellent resource for
studying stellar variability.

\section{Variability studies with SuperWASP}

\begin{figure}[h]
\begin{center}
 \includegraphics[width={0.49\textwidth}]{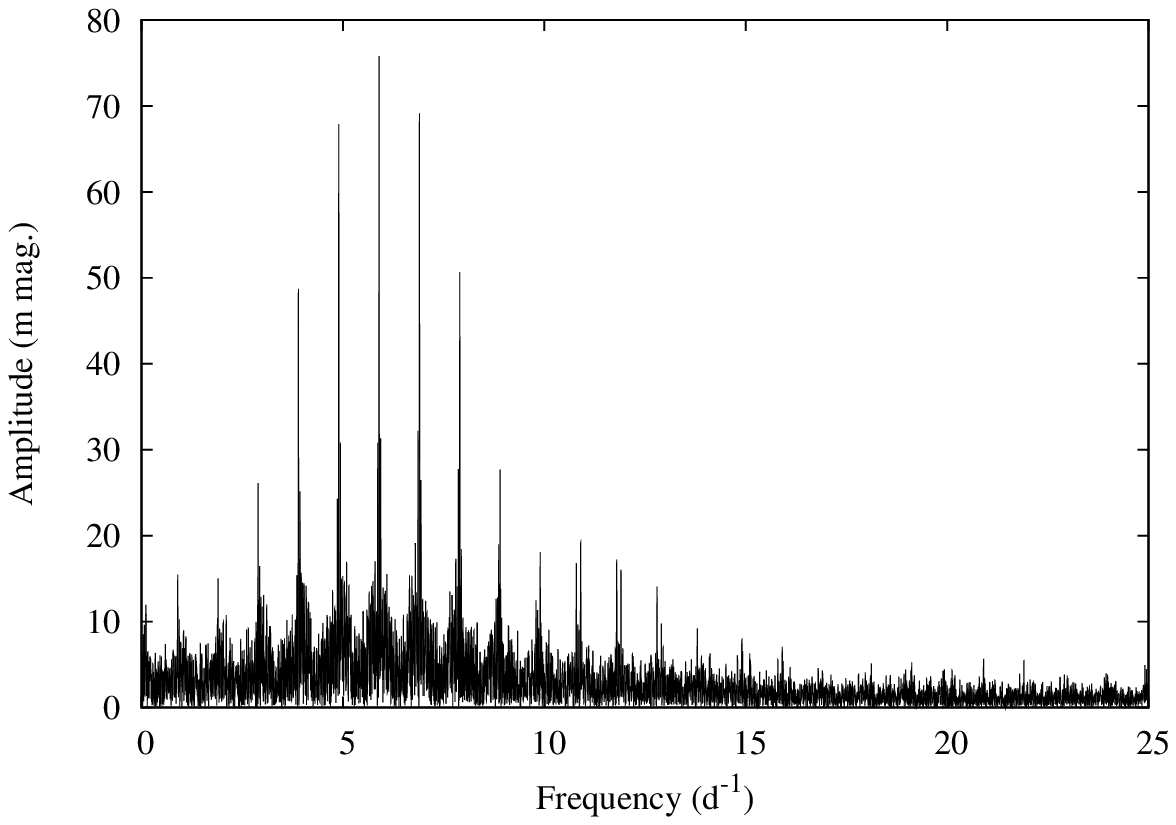}
 \includegraphics[width={0.49\textwidth}]{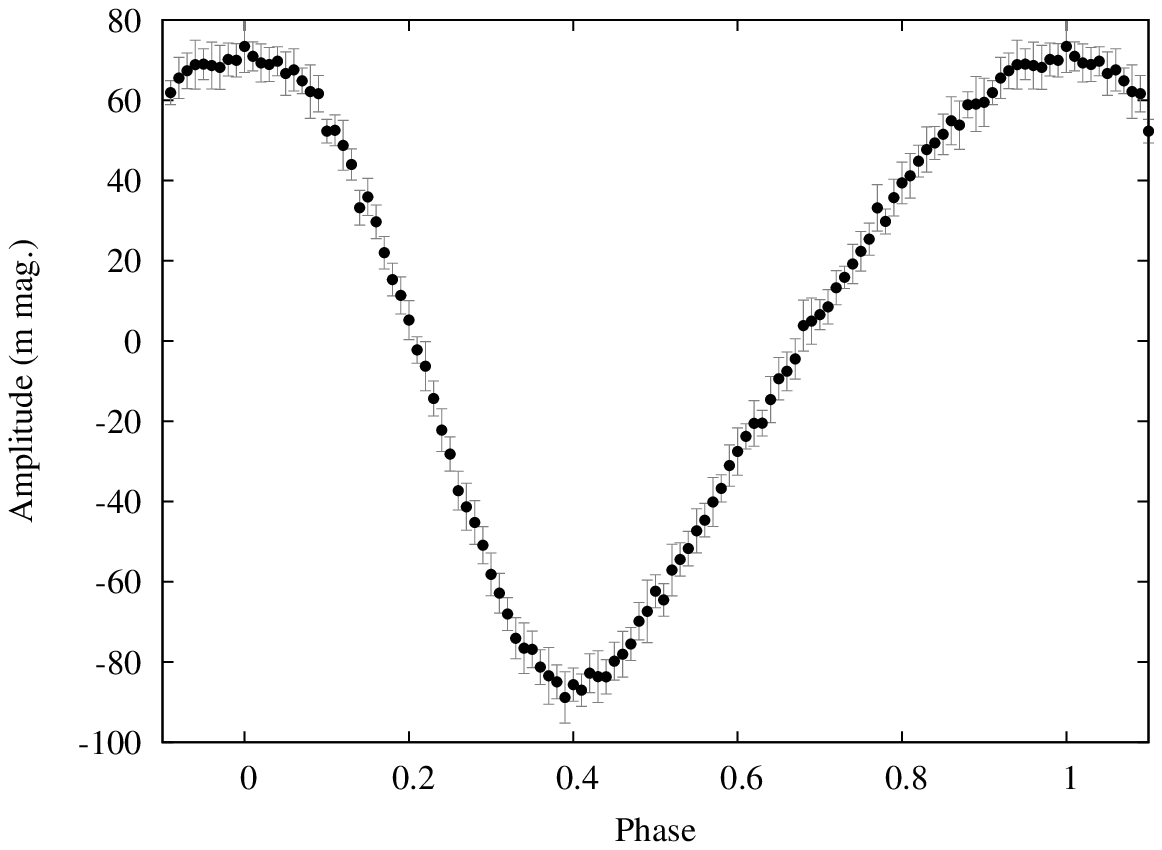}
 \caption{WASP periodogram (left) and lightcurve folded on the 0.17-day
 period (right) of the $\delta$ Sct variable UY Col (HD\,40765).}
 \label{UYCol}
\end{center}
\end{figure}

There are many stars in the WASP archive that exhibit high amplitude pulsations
over different timescale from hours to several days, including numerous RR
Lyrae, Cepheid and $\delta$~Scuti variables (Fig.~\ref{UYCol}). This is, of
course, in addition to many eclipsing and short-period binary systems (e.g.
\cite[Norton et al. 2013]{2011A&A...528A..90N}). By cross-matching WASP
photometry with high-precision lightcurves from {\it Kepler}, Holdsworth (PhD
Thesis, in prep.) found a pulsation detection limit of 0.5 m\,mag. is possible
for stars brighter than mag. 10.

\subsection{Pulsations in Am stars}

The large sky coverage of the SuperWASP data allows for statistical studies of
the pulsational properties of selected groups of stars. In an extensive study,
\cite[Smalley et al. (2011)]{2011A&A...535A...3S} took the \cite[Renson \&
Manfroid (2009)]{2009A&A...498..961R} catalogue of Am stars and selected stars
with greater than 1000 WASP data points. Excluding eclipsing binaries, 1620 Am
stars were studied using Lomb periodograms to select candidate pulsating stars.
These were subsequently examined in more detail using {\sc period04} \cite[(Lenz
\& Breger 2005)]{2005CoAst.146...53L}. A total of 227 pulsating Am stars were
identified, representing 14\% of the sample. Pulsations in Am stars are more
common than previously thought, but not where expected (\cite[Turcotte et al.
2000]{2000A&A...360..603T}).

While the amplitudes are generally low, the presence of pulsation in Am stars
places a strong constraint on atmospheric convection, and may require the
pulsation to be laminar.  While some pulsating Am stars had been previously
found to be $\delta$~Sct stars, the vast majority of Am stars known to pulsate
have been found by SuperWASP, thus forming the basis of future statistical
studies of pulsation in the presence of atomic diffusion.

\subsection{Rotational modulation in Ap stars}

WASP photometry can be used to investigate the rotational modulation of stars,
as is routinely performed as part of the planet detection programme
(\cite[Maxted et al. 2011]{2011PASP..123..547M}). Rotational modulation is also
be seen in the magnetic Ap stars. For example, WASP photometry was used, in
conjunction with that from ASAS and Hipparcos, to improve the rotational
ephemeris for HD\,96237 (\cite[Elkin et al. 2011]{2011MNRAS.411..978E}).

\subsection{Rapidly Oscillating Ap stars}

\begin{figure}[h]
\begin{center}
 \includegraphics[width={0.49\textwidth}]{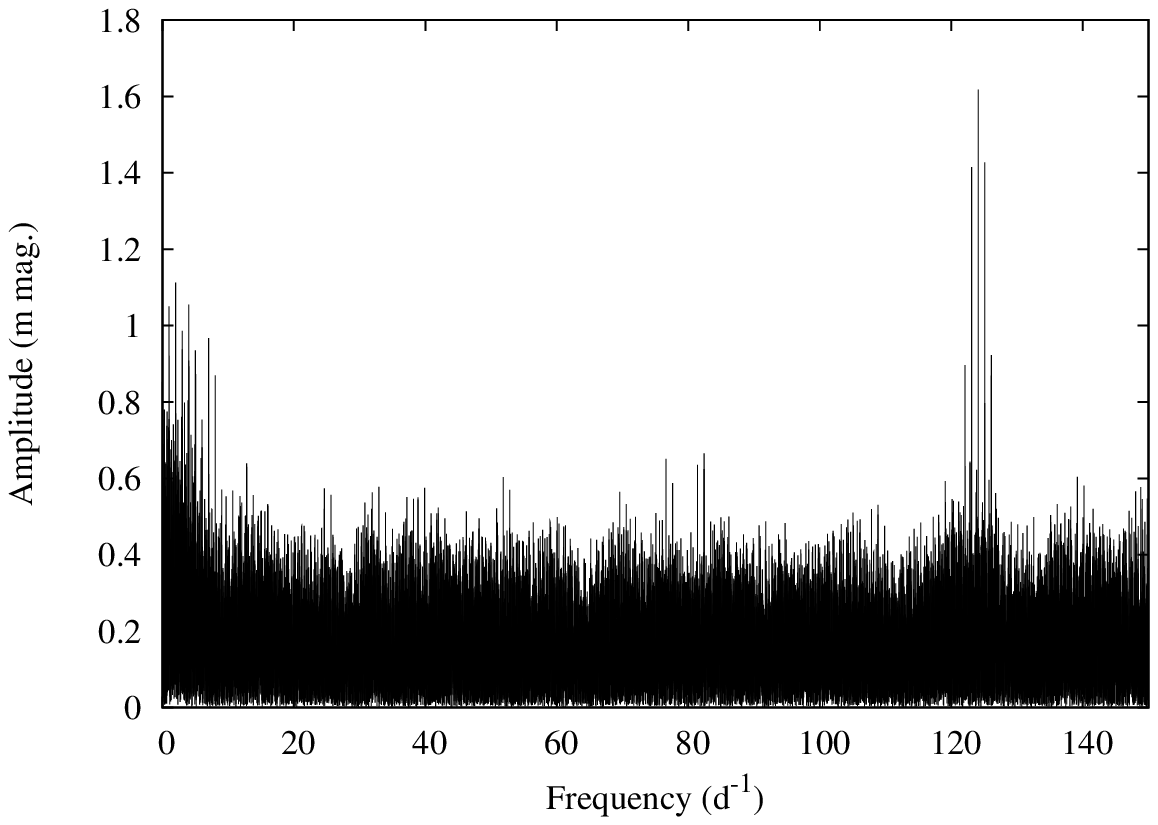}
 \includegraphics[width={0.49\textwidth}]{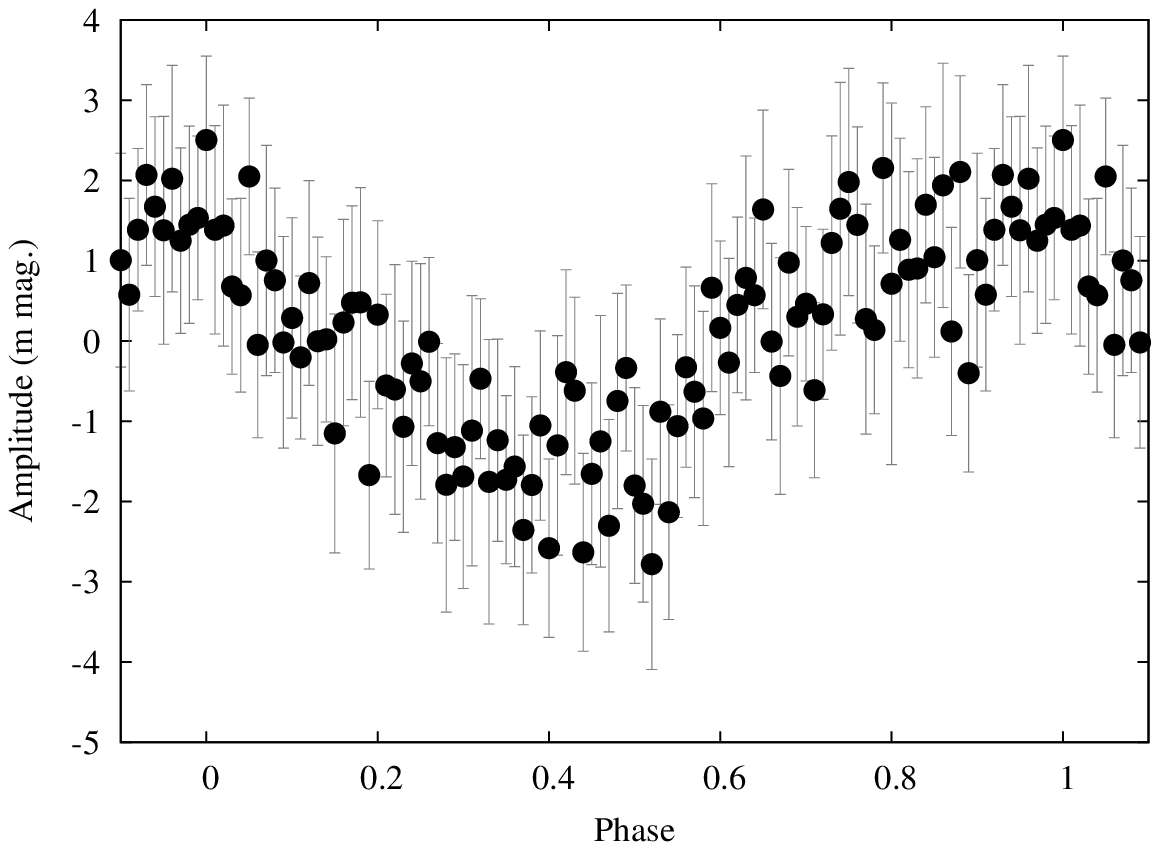}
 \caption{WASP periodogram (left) and lightcurve folded on the 11.6-minute
 period (right) of the roAp star BN Cet (HD\,12932).}
 \label{BNCet}
\end{center}
\end{figure}

The rapidly oscillating Ap (roAp) stars are a relatively rare subset of the
magnetic Ap stars which exhibit short period oscillations. There exists
lightcurves of several known roAp stars in the WASP archive, including HD\,12932
(BN Cet). This star exhibits 124.1\,d$^{-1}$ (11.6 min) oscillations
(\cite[Schneider \& Weiss 1990]{1990IBVS.3520....1S}) and are clearly detected
in the WASP periodogram (Fig.~\ref{BNCet}).  The amplitude in the WASP filter
is approximately half that in $B$, consistent with the expected variation of
amplitude with filter wavelength (\cite[Medupe \& Kurtz
1998]{1998MNRAS.299..371M}).

For stars with sufficient WASP photometry, no new roAp stars have been found
among the Ap stars in the \cite[Renson \& Manfroid (2009)]{2009A&A...498..961R}
catalogue. Nevertheless, a systematic search of over 1.5 million A and F type
stars in the WASP archive has yielded over 200 stars with pulsation frequencies
higher than 50\,d$^{-1}$. Subsequent spectroscopic follow-up has confirmed that
at least ten of these stars are new roAp stars (\cite[Holdsworth \& Smalley
2013]{HoldsworthSmalley}).

\subsection{Ultra-high frequencies}

\begin{figure}[h]
\begin{center}
 \includegraphics[width={0.49\textwidth}]{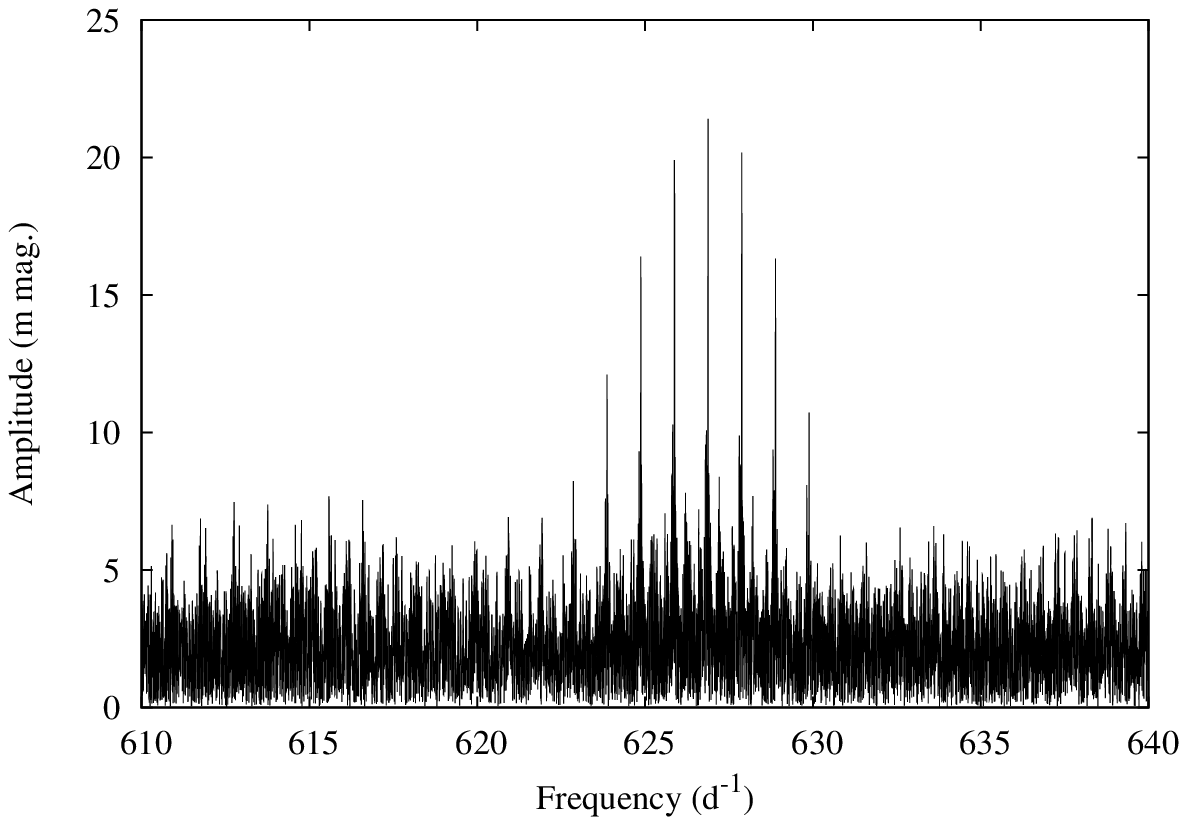}
 \includegraphics[width={0.49\textwidth}]{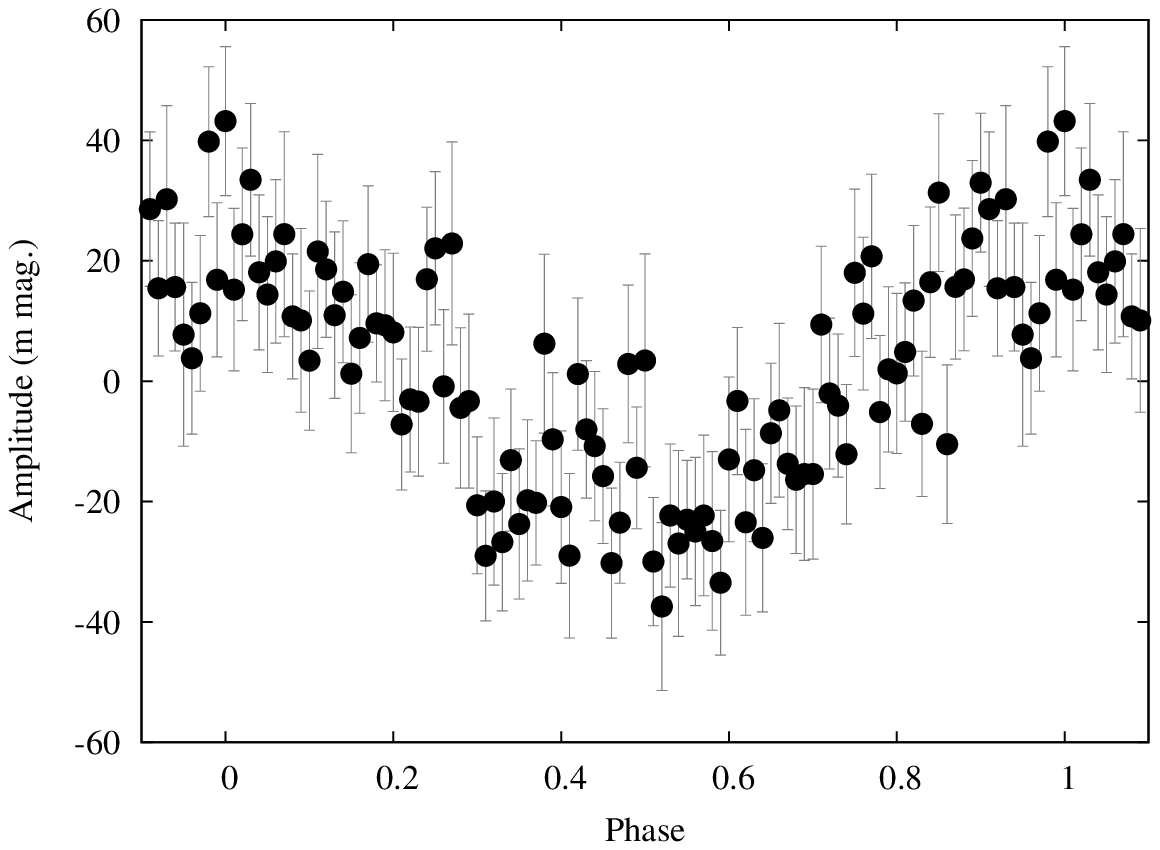}
 \caption{WASP periodogram (left) and lightcurve folded on the
 2.3-minute period (right) of the sdB pulsator QQ Vir.}
 \label{QQVir}
\end{center}
\end{figure}

Pushing to higher frequencies, WASP photometry can be used to search for
frequencies up to $\sim$1000\,d$^{-1}$. For example, the sdB star QQ Vir has
626\,d$^{-1}$ (2.3 min) pulsations (\cite[Silvotti et al.
2002]{2002A&A...383..239S}). Figure~\ref{QQVir} shows the WASP periodogram where
the pulsations are present with an amplitude of 0.02 mag. A preliminary search
of the WASP archive for very rapidly pulsating stars has resulted in the
identification of a hot sdBV star with 636\,d$^{-1}$ pulsations and an amplitude
of only 8~m\,mag.

\subsection{Signal in noise}

\begin{figure}[h]
\begin{center}
 \includegraphics[width={0.49\textwidth}]{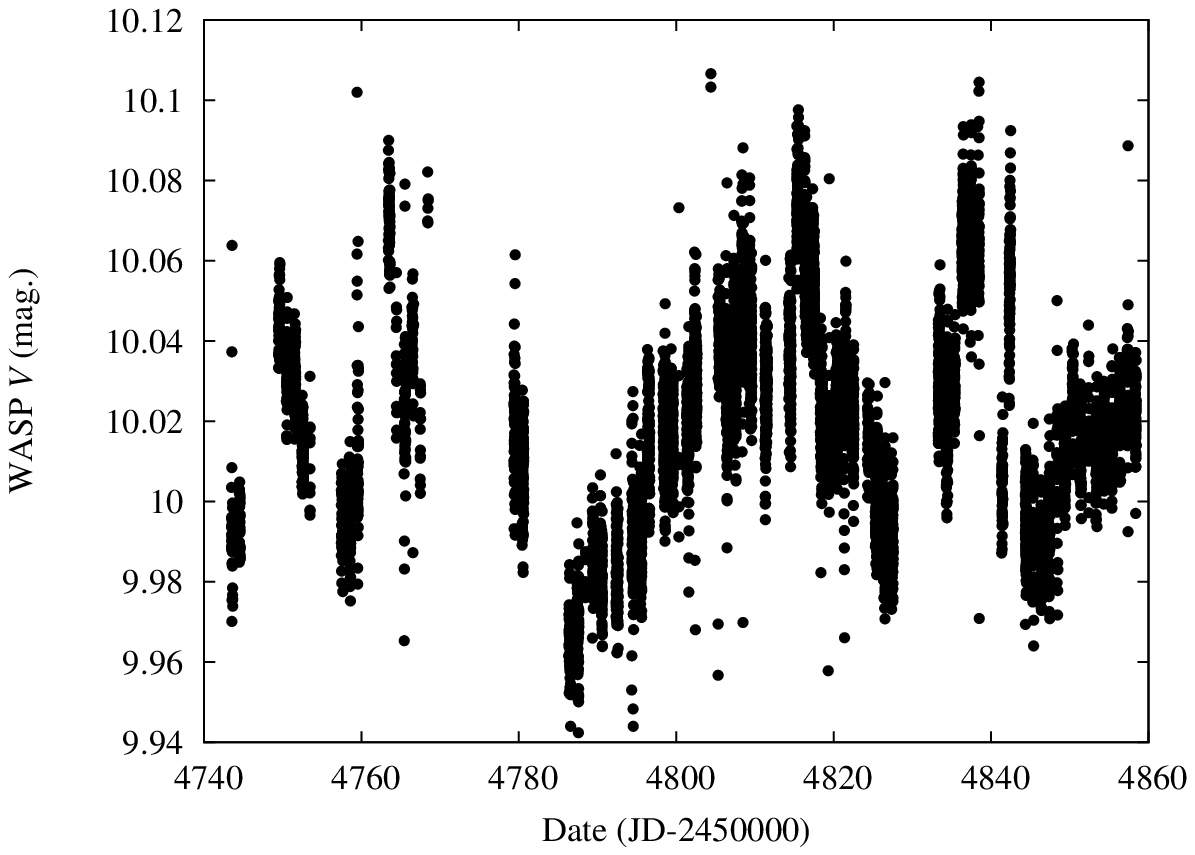}
 \includegraphics[width={0.49\textwidth}]{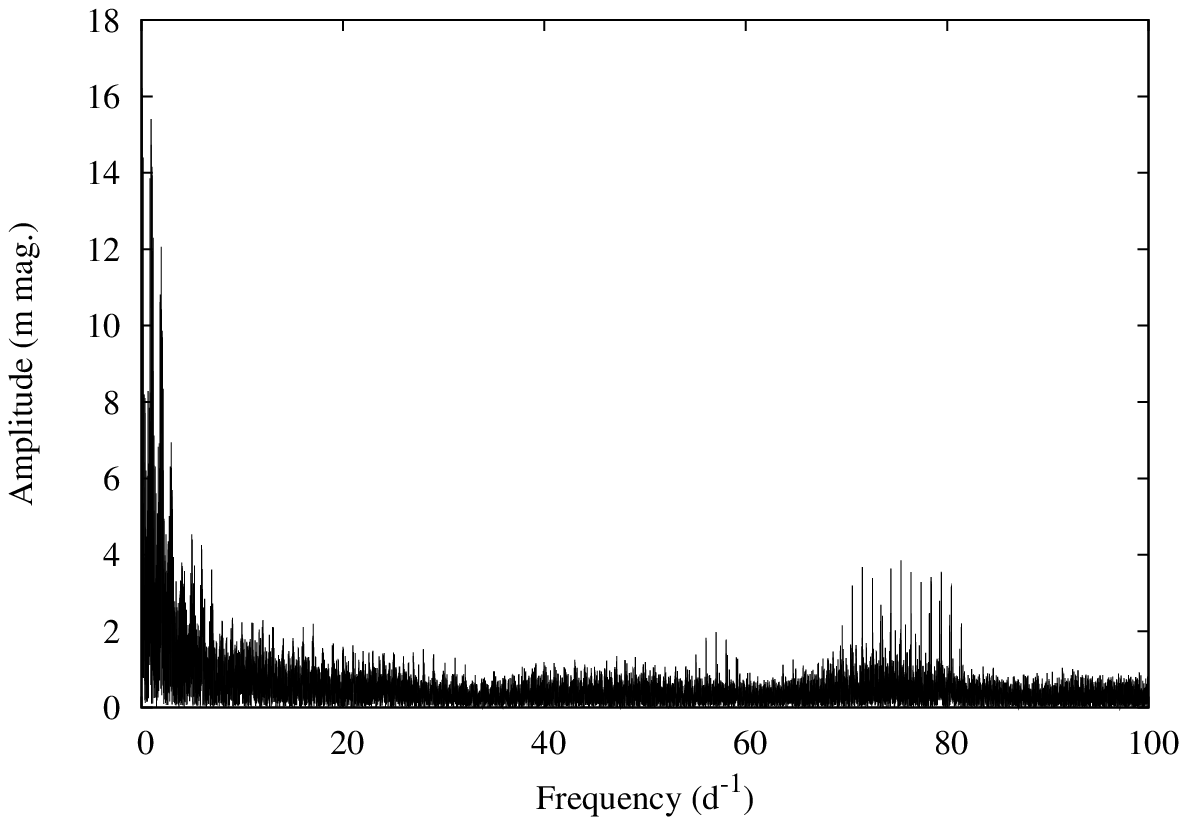}
 \caption{WASP lightcurve (left) and periodogram (right) of the
 pre-main sequence star V1366 Ori (HD\,34282) showing random variations due to
 dust, but with periodic short period $\delta$~Scuti-type pulsations.}
 \label{V1366Ori}
\end{center}
\end{figure}

During the WASP project's trawl for planetary transits, lightcurves are
encountered which appear very poor and unsuitable for planet hunting. However,
closer inspection reveals that some are actually due to real stellar
variability. For example, HD\,34282 (V1366 Ori) has an apparently random noise
lightcurve (Fig.~\ref{V1366Ori}) but the periodogram recovers the 79.5\,d$^{-1}$
and 71.3\,d$^{-1}$ $\delta$~Scuti-type pulsations known to be hiding in the
dusty environment of this pre-main sequence star (\cite[Amado et al.
2004]{2004MNRAS.352L..11A}).

\section{Summary}

The WASP archive contains broadband photometry for over 31 million stars with a
precision of $<$0.01 mag. The observing strategy yields two consecutive
30-second exposures every 10 minutes. The relatively large 14$''$ pixels and
48$''$ photometry aperture means that blending and dilution can be an issue for
certain stars. Nevertheless, the photometry has proved invaluable in the
investigation of the statistical occurrence of variability and allowed for the
identification of new members of several classes of variable stars. There are
certainly a lot of interesting stars in the WASP archive.

\end{document}